# Enabling Value Sensitive AI Systems through Participatory Design Fictions

**Q. Vera Liao and Michael Muller**

IBM Research
vera.liao@ibm.com, michael_muller@us.ibm.com

**Abstract**

Two general routes have been followed to develop artificial agents that are sensitive to human values---a top-down approach to encode values into the agents, and a bottom-up approach to learn from human actions, whether from real-world interactions or stories. Although both approaches have made exciting scientific progress, they may face challenges when applied to the current development practices of AI systems, which require the understanding of the specific domains and specific stakeholders involved. In this work, we bring together perspectives from the human-computer interaction (HCI) community, where designing technologies sensitive to user values has been a longstanding focus. We highlight several well-established areas focusing on developing empirical methods for inquiring user values. Based on these methods, we propose *participatory design fictions* to study user values involved in AI systems and present preliminary results from a case study. With this paper, we invite the consideration of user-centered value inquiry and value learning.

## Introduction

With the rapid development of AI technologies, there is a growing interest in both the scientific community and the public to embed ethics and human values in autonomous agents (Russell et al. 2017). A main thread of research explored adapting decision frameworks by embedding values as rules, constraints or preference structures (Rossi 2015; Greene et al. 2016). Most AI researchers have focused on the computational implementation of these models and delegated the specification of values to classic ethical theories such as utilitarianism and deontology. However, to allow AI systems to work in practice, one may need to specify values or constraints that are consistent with the values of actual users and their communities (Cheon and Su 2016). Considering the complexities of human value systems (Schwartz 2012) – even for people in a shared context (Draper et al. 2014) – there are many arguments why being bounded by or optimizing for some utilitarian values such as happiness or welfare may fail, or rapidly become obsolete, for the near-term AI technologies working in narrow domains.

To complement the top-down approach, which may fail to account for nuanced choices with rigid specification, some researchers explored bottom-up approaches to learn the tacit values from human actions (whether observed or interacted with) (Hadfield-Menell et al. 2016; Riedl and Harrison 2016). Naturally, such approaches would require large amount of human choices data. A promising direction that AI researchers have started to explore is learning from "crowdsourced" choices with story scenarios, where the story setup constrains the action space. For instance, the "trolley problem" is a widely used scenario of ethical dilemma, for which large crowdsourced datasets have been created in the hope of informing the development of "moral machines" (Bonnefon et al. 2016; Shariff et al. 2017). However, before we can map out a complete picture of human activities, such approaches are not ready to relieve us from the needs to understand values in specific domains and of specific stakeholders that are directly or indirectly impacted by the technology (Friedman et al. 2013).

To understand the values of stakeholders, task domains and different user communities is, in fact, a core undertaking of the HCI community, with the dedicated research area of *value sensitive design (VSD)* (Friedman 1996; Friedman et al. 2013). The tension between the top-down approach by seeking normative, utilitarian values and the bottom-up approach by empirically inquiring user values is also a familiar one to the HCI community (Borning and Muller 2012). In the following section, we will review this line of work and methods in *participatory design* (PD) to study diverse values of stakeholders. We also introduce design fictions (DF) as a *speculative design* method to understand user needs for technologies that do not yet exist – i.e., emerging or near-future technologies. Based on these approaches, we propose a set of user engagements with *participatory design fictions*

(PDF)---soliciting stakeholder input with strategically incomplete fictions---to analyze the value issues, including but not limited to types of values and their priority orders, for AI applications.

We consider PDF to potentially complement both the top-down and bottom-up approaches for computationally embedding human values in AI models. While it is straightforward to encode the value issues identified from analysis of PDF as constraints or preference structures, we also see a path to data-driven approaches by iteratively creating more optimal scenarios to collect human decision data for the agents to learn from. That is, instead of attempting to generalize machine moral decisions from a single scenario such as the trolley problem, we seek to create more application-specific scenarios to make the value learning task more tractable, and more applicable. We would like to draw a parallel with *user-centered design* in conventional technology development process to invite the consideration of *user-centered model development* for AI technologies to be more sensitive to the values of its diverse stakeholders. Just like user-centered design, this process can be iterative, moving from open-ended discursive fictional spaces for formative understanding of the scope of values, to more well-defined scenarios (like the trolley problem) for quantitatively validating and identifying nuanced value issues.

## HCI Approaches to Values

We start by introducing three established HCI areas that are concerned with studying and designing for user values. We note that each scholarship tradition has long threads of theories and methods that go beyond what are reviewed here, where our primary focus is on value inquiry and their pragmatic use for AI technologies.

### *Value Sensitive Deign*

Originally introduced by Batya Friedman (1996), value sensitive design (VSD) is "*a theoretically grounded approach to the design of technology that accounts for human values in a principled and comprehensive manner throughout the design process*". VSD was developed in an effort to consider human values with ethical import as a central technology design criterion, along with traditional design criteria such as performance and usability. Freedman et al. developed a "tripartite methodology", consisting of iterations of conceptual, empirical and technical investigations. Specifically, conceptual investigations are concerned with careful conceptualization of fundamental value issues raised by the technology at hand, asking questions such as "*what values and whose values should be supported by the design process?*" Empirical investigations further complement the understanding through investigations of the human activities in which the technical artifact is situated, asking more fine-grained questions such as "*how do stakeholders apprehend values in the context? How do they prioritize different values?*" Lastly, the technical investigation considers how technological properties and alternatives (e.g. different designs) support or hinder human values identified in the conceptual and empirical analysis. For the scope of these investigation, VSD emphasizes the involvement of both *direct stakeholders* (direct users) and *indirect stakeholders* (who may be indirectly impacted).

In its essence, VSD centers upon the explication of value issues impacting diverse stakeholders of a technology and embedding them as goals and constraints in the design process. This shares much of the same spirit with the currently pursued approaches for embedding human values in AI models. In an early paper, VSD recommended 12 human values that hinge on ethical theories including human welfare, ownership, autonomy, privacy, etc (Friedman and Kahn 2003). But critique ensued, including a lack of ground to proclaim normative values with complex technologies and ever-changing social contexts (Manders-Huits 2011), and biases towards western, liberal and civic values (Borning & Muller 2012). More recently, a set of empirical innovations emerged within the VSD framework on soliciting values oriented considerations from stakeholders using interviews, card-based exercises, etc. (e.g., Friedman and Hendry 2012; Woelfer et al. 2011). Our work in what we call *value sensitive inquiry* is built on this foundation.

### *Participatory Design*

Participatory design (PD) is a set of theories and practices to bring end-users as full participants into activities leading to technological products (Bodker 2014). A main focus of PD research is the development of diverse tools and techniques to bridge participants' current practices and the abstract design space of new or near-future technologies, and enable fruitful dialogues between developers/researchers and end users (Kensing and Blomberg 1998; Muller and Druin 2010). Narrative structures play a key role in serving these goals by avoiding abstract representation, and allowing end users and designers to experiment with various design possibilities (Kensing and Blomberg, 1998). Narratives in PD have taken various forms including scenarios, story-telling (e.g. co-creation of photograph and drama), game playing (e.g. card games) and co-creation of prototypes (Muller and Druin 2010).

In the next section, we will consider adapting tools and techniques from PD researchers' repertoire for value inquiry or value learning. The bottom-up approaches of learning values from human choices in scenarios of ethical dilemma may be considered a distributed form of participatory design, although PD have traditionally adopted more qualitative analysis on the narratives. We note that PD owes its roots in value issues with the political movement in the 1970s to entitle workers along with the management to determine what technologies to bring into the workplace

(Winograd and Kuhn, 1996). PD thus embodies moral commitments to diverse human values, meanwhile a more descriptive, bottom-up approach to integrate them in technology design. PD potentially offers solutions to the recent warnings raised by researchers of a techno-centric view to AI system design, which puts users and the society in a passive role of accepting and adapting to these technologies (Sabanovic 2010; Cheon and Su 2016)

*Design Fictions*

A unique issue in considering values involved in AI systems, at least in the current term, is a future orientation. This is not only because many progressing AI technologies are not yet a part of ordinary people's everyday life (e.g., self-driving cars, humanoid robots), but also because of the rich depiction of AI technologies in popular culture's imagined futures (e.g. science fiction books and movies). Although some argue that the pre-exposure to fictional experience may implicate challenges in the design of AI technologies, they may in fact provide an experimental space for values issues, which are considered more abstract and relatively stable constructs.

This idea converges with the recent development of Design Fictions (DF), a genre of speculative design methods, as first introduced by Sterling's definition of "deliberate use of diegetic prototypes to suspend disbelief about change" (2012). In particular, DF has been embraced for revealing values associated with new technologies (Dourish and Bell 2014; Tanenbaum et al. 2016); for opening a space of diverse and polyvocal speculations about future technologies (Blythe 2014); and engaging and inquiring about specifications of these values (Draper et al. 2014). Compared to conventionally used scenario-based design methods, DF focuses more on constructing an engaging fictional space for speculating and provoking, some calling it "fieldwork of the future" (Odom et al. 2012).

We introduce fictional narratives to explore the value issues of AI technologies as a natural extension of the fictional world people have already been exposed to through popular culture. Importantly, we argue for carefully constructed fictions that put the targeted technology and its personal and social contexts in the focal point. It is necessary to construct technology-specific fictions because machine ethics may be a complicated issue and it may not be sufficient to generalize from human ethical standards. For example, in a recent *Science* paper, through a survey study, researchers found that the trolley problem is further complicated with the replacement of self-driving car for the human controlling the lever (Bonnefon et al. 2016).

## Participatory Design Fictions (PDF): Towards Value Inquiry for AI Technologies

In summary, we introduce perspectives from the HCI communities to begin to think about empirical methods to enable value inquiry and value learning for AI applications. We employ the stakeholder concepts and an iterative combination of conceptual and empirical investigation of values from Values Sensitive Design. We borrow tools from Design Fictions to elicit value issues involved in future-oriented AI technologies. Finally, we use Participatory Design methods to de-center the inquiry from technology experts, and to re-center the inquiry on the users.

In this section, we consider some examples of participatory design techniques and adapt them for introducing *fictional experience spaces* where potential stakeholders can individually or collectively envision, design, and critique emerging or future AI technologies. We expect value issues to be articulated in the discursive story world in a structured way, as enforced by the carefully constructed story boundaries.

Ultimately, this research area should strive to provide a set of guidelines for constructing such boundaries and enabling effective value inquiry. Meanwhile, we also advocate an iterative process to start with more formative inquiry with open-ended discursive spaces (e.g., providing just a beginning of a story and main characters), asking questions such as "*what values and whose values should be considered?*". Then one can refine the storyline to narrow down to more analytical inquiry of the key value-sensitive system actions identified, asking fine-grained questions such as "*how to resolve the diverse or conflicting values in this situation?*" Innovative setup that allows social emergence of critical value issues such as collective deliberation platforms and social games can be adopted. It is also possible to work iteratively in parallel with the AI model development process, for which the fictional experience and scenario-based model testing can be engaged in at the same time.

**Fictions as Probes**. A main usage of stories in PD is to elicit users' evaluation as probes in user interviews or focus groups (e.g., Sato and Salvador 1999). In this way, we could present fictions regarding the targeted technology as starting points for a conversation. Questions addressed to users could focus on the support or hindrance of values that they perceive in the stories. For example, Draper et al. (2014) used fictional probes to elicit ethical evaluation of eldercare robots with focus groups. Cheon and Su (2016) used short fictions as evaluation cases to solicit roboticists' considera-tions of value issues in their development work. This approach can be considered a more close-ended employment of fictions for value inquiry.

**Fictions as Conceptual or Literal Guerrilla Theatre.** Soliciting tacit value issues is a challenging task and PD researchers have pursued more critical approaches by inviting users to change the stories. Boal's Theatre of the Oppressed employed stories (enacted as brief dramas, portrayed in the street) as a critique of power (1992). Surprised passers-by were encouraged to rewrite the story – or to participate in its changed enactment - so that it would have an outcome that they preferred. PD researchers derived similar theatre approaches as a means to critique and change a proposed user experience (Muller et al. 1994). For work with future-oriented AI technologies, plot devices, whether in written or performative forms, can be adopted to invite users to engage in value (re) configurations by rewriting of characters, plots and outcomes.

**Fictions as Participatory Constructions.** A more ambitious approach in PD is to solicit stories to be written by users. For instance, Beeson and Miskelly (2000) advocate user-created stories through hypermedia technologies such that "plurality, dissent, and moral space can be preserved." Druin engages children to construct their own narratives in a playful technology environment (2002). Prost et al. described a structured, participatory workshop process to teach students to create design fictions for sustainability technologies (2015). We envision engaging users in completing their own versions of stories by posing value-oriented questions as starting points. For example, a fictional character facing a complex choice would require explicit consideration and articulation of values, and also encourage fictional simulation of outcomes with different "value configurations". In the next section, we present a preliminary case study experimenting with this approach.

**Fictions as Group Co-Creations**. There are two reasons to consider collective creation of fictions. One is a possibility for social emergence of critical value issues, by preserving and building upon social traces of others. In a recent large-scale online experiment to collect moral decision data for the trolley problem, MIT researchers introduced the mechanism of crowd-created scenarios to be judged by others, who could be inspired to produce variations that are centered around core value issues (Shariff et al. 2017). By making articulated values visible and actionable to other participants, one can thus enable a collaborative process that makes critical value issues emergent.

Another reason to pursue group co-creation is that stakeholders may also interact with AI technologies as groups or in social contexts, so we need ways to collectively speculate about these technologies. In PD, photodocumentaries have been co-created by communities. For example, to address the problem that "rural women are often neither seen nor heard," Wang et al. (1996) invited Chinese village women to articulate their lives through photo novellas with the goal of influencing policy-makers. We could leverage digital technologies to enable group co-creation of values-oriented stories about AI technologies. For example, online games have been recently considered as collective venues for design fictions (Tanenbaum et al. 2012).

*Data from PDF for AI Models: Possible Directions*

The user narratives collected from PDF should be considered data. Analyses on the data can generate high-level insights for value issues around a targeted AI technology, and can also be used as direct input for learning based models. Following the tradition of inductive content analysis in social science research, Grounded Theory based analysis can be used to identify value issues in the collected narratives. Grounded Theory provides a set of rigorous manual coding procedures leading to the emergence of *conceptual categories* (Charmaz 2016). Recently, researchers started exploring the parallel between unsupervised learning for pattern discovery and Grounded Theory based manual analysis on relatively "small data" (Muller et al. 2016; Baumer et al. 2017). We consider the complementariness of qualitative value inquiry and data-driven value learning to be a promising future research area for analyzing narratives collected from PDF.

Another area to explore is to complement the reinforcement learning based frameworks for AI ethics, for example, to fill the gaps in specifying utility or reward functions based on empirical investigation. The idea of inverse reinforcement learning (IRL) by modeling the reward function as a probabilistic inference problem by observing human actions or feedback is particularly interesting. If one can map the agent actions, and even the learned reward functions, to a fictional space, we can envision many possibilities of eliciting feedback and for these "real stories" of AI agents under ethical development. This is in line with the recent proposal of cooperative inverse reinforcement learning, and leveraging active learning, active teaching and communicative actions, to achieve effective value alignment of AI systems (Hadfield-Menell et al. 2016).

**PDF: A Case Study**

As a first step, we have conducted multiple experiments with using incomplete design fictions to engage users in writing their own stories about future applications of AI applications. These experiments aimed to explore how future users react to the probe of strategically incomplete stories and the scope of generated narratives. By design, each fiction begins with a setting, one or more characters, and a plot. The fiction contains brief *diegetic aspects*, such as suggest-ing concerns and motivations of the protagonist (s). The stories centers around the use of the *targeted AI technology* and end prematurely at an unresolved *decision point*. We invite the participants to complete the story, or to provide details that are implied by the story (e.g., "what choice do you think the character should make?" or "how would you design the robot?"). By posing value-oriented questions with a rich but bounded choice space, we hope to elicit narratives that can articulate values explicitly or implicitly involved in using the technology. Below, we present one example of such an incomplete story we used:

### The Nanny Bot

> We had both taken parental leave with Ariel, our new baby, but now we needed to return to full-time work. Because our work schedules were unpredictable, we needed a part-time, on-request nanny to cover an extra 1-2 hours at the end of the day.
> So of course we went to the Bot Store website, where we could order and personalize our nanny bot. There were so many choices! Our close friends had made their nanny specs viewable. We copied their nanny specs as our starting point, and then we began to personalize those specs to suit our family. We examined some of the attributes that we could choose, like personality, skin color, gender……
>
> [complete the story here]
>
> When we'd finished the choices that were important to us, we accepted the other personalization from our friends' records. And then we were done! We were excited to be bringing the nanny-bot into our lives, and into Ariel's young life.

This incomplete story involves a targeted AI technology in a specific domain (childcare). The setup is a more distant future than the current technological capabilities to allow broader value inquiries. The story is left blank at the shopping decision point, a familiar decision context that both serves to reduce abstraction of the technological space and imposes a need to explicate values. In the last paragraph, we mention the consideration of "friends' records" to nudge for social considerations (i.e., "what I value in relation to what others value"). The brief story ends at a prospective point that makes further exploration and simulating the outcomes of the decision configuration possible.

In our preliminary exploration, we experimented with different formats of inquiries. To date, we engaged six individual volunteers for interviews and writing tasks, and a short group design workshop at the Human Computer Interaction Consortium (HCIC) 2017. With the interview format, the researcher could actively engage in interactive inquiries, and we found it effective in soliciting the *reasoning* behind the choices, which in itself provides value relevant information, including priority orders and conflicts of values. In the writing task format, participants were more willing to engage in exploring the *outcomes* of value choices by writing lengthier "what happens afterwards" storyline. To preserve the integrity of the participants' narratives, the inquiry protocol for writing formats should be kept short and general (e.g., "what decision would you make?") instead of asking multiple progressive questions as in interviews. In the design workshop, we adapted the task into an "artifact design" task. Instead of asking about the shopping choices of the nanny bot, which may be difficult for the group to resolve conflicting specifications of individual members, we asked the group to create a fictional inventory of the bot shopping website (i.e., what bot features can be specified), which would suggest collective "pooling" of values.

Following grounded-theory based analysis, we identified a set of values that can be embedded in developing childcare AI systems. Some of the values are among the values identified to have universal "ethical import" in Friedman et al.'s VSD papers (2003; 2012), including accountability (e.g., "*punctual*", "*never have accidents*"), trust, and identity ("*how I would interact with the baby…our friend's values may be different*"). Some are less obvious and specific to the childcare context, such as family tradition (e.g., language, appearance and habits) and social development for the child (e.g. "*quality time with the baby*", "*encourage interaction with others*") We also note that certain values are not universal or held to the same standard. For example, participants expressed highly varied requirements for robot autonomy and privacy.

Results collected from the small-scale studies are encouraging, demonstrating the potentials of using strategically designed incomplete fictions to elicit rich and complex value issues, especially ones that might be ignored if we took, instead, a purely normative view on values. Meanwhile, the observation suggests that informants' reactions to PDF are highly diverse and individualized. On the one hand, this implies challenges for developing principled methods to construct the fictions, as well as for data-driven approaches to discover values from narratives. On the other hand, the observations further highlight the needs to start from more open-ended formative inquiry without presuming values or even the narrative frameworks. In some of our encounters, participants questioned our conceptualization of opposing values to begin with. For example, in another incomplete fiction we experimented with inquiring about labor issues, by asking participants how the owner of an apple orchard should make choices between robot apple pickers and migrant workers. Participants pushed back the binary choice and suggested creative solutions that imply more complex value configurations (e.g., "*employ human workers for the most expensive apple varieties…and robots for the others*").

Lastly, we consider the possibilities of extending the fictional setup beyond a written or conversational format, and beyond individual creations. For example, the experience of "bot shopping" and "bot store" can be readily developed into performative actions and spaces for more immersive user engagement that can be participated by groups (similar to the User Enactments, physical objects and spaces used by Odom et al. to investigate designs of radical technologies (2012)). The possibilities for digital games are particularly promising for gathering large-scale narratives data and social emergence of value issues. We note that popular simulation games such as SIMS already embody some of the key elements such as the diegetic aspects, decision point and bounded space. By employing game plots with carefully constructed value prompts, not only can one elicit more natural values-oriented narratives, but also it is possible to enable large-scale collective co-creation of "possible, probable and preferred future", providing empirically and socially validated rich values data for AI systems to evolve their moral status.

## Conclusion

The goal to embed human values in AI system is critical and inevitable. As the scientific community develops computational solutions, so should we make progress in innovating empirical methods to elicit and understand values of different user communities and in different tasks (IEEE 2016). This work represents an effort in that direction by proposing Participatory Design Fictions. We build our foundation on the basis of three established HCI areas studying user value issues, and consider adapting existing user research methods to: 1) enable the explication of broad and tacit value issues around future-oriented technologies by leveraging fictional spaces and diegetic elements; 2) more closely serve value inquiry and value learning by leveraging strategically incomplete stories that centers upon targeted technology and critical decision points. We present preliminary insights from our case studies and invite joint effort across scientific communities for user-centered approaches to value inquiry and value learning for AI systems.